\documentstyle[preprint,aps,prc,epsfig]{revtex}
\tightenlines
\newenvironment{lemma}{{\bf Lemma:} \noindent \it} {}
\newenvironment{proof}{{\bf Proof:} \noindent}
                       {\begin{flushright} {\em End of proof.} 
                        \end{flushright}}
\begin{document}
 
\begin{titlepage}
\title{A discrete Hubbard-Stratonovich decomposition for general,
       fermionic two-body interactions}
\author{S. Rombouts \thanks{Postdoctoral Fellow of the Fund for Scientific 
                            Research - Flanders (Belgium)},
        K. Heyde and N. Jachowicz}
\address{Vakgroep Subatomaire en Stralingsfysica
         \\
         Institute for Theoretical Physics
         \\
         Proeftuinstraat 86, B-9000 Gent, Belgium
         \\
         tel: \#32/9/264.65.41 fax: \#32/9/264.66.99
         \\
         E-mail: Stefan.Rombouts@rug.ac.be, Kris.Heyde@rug.ac.be
         }
\date{\today}
\maketitle

\begin{abstract}
\noindent
A scheme is presented to decompose the exponential of a two-body operator
in a discrete sum over exponentials of one-body operators. 
This discrete decomposition can be used instead of the 
Hubbard-Stratonovich transformation in auxiliary-field 
quantum Monte-Carlo methods.
As an illustration, the decomposition is applied to the Hubbard model, 
where it is equivalent to the discrete Hubbard-Stratonovich transformation
introduced by Hirsch, and to the nuclear pairing Hamiltonian.
\end{abstract}


\end{titlepage}
\newpage

\section{introduction}
In auxiliary-field quantum Monte-Carlo methods (AFQMC), 
such as the projector, grand-canonical \cite{lind} and 
shell-model quantum Monte-Carlo methods \cite{smmc}, 
the Boltzmann operator $e^{-\beta \hat{h}}$, with $\hat{h}$ the Hamiltonian,
is decomposed in a sum or integral of exponentials of one-body operators.
This sum or integral is then evaluated using Monte-Carlo techniques.
For the decomposition, these methods rely on the Hubbard-Stratonovich
transformation \cite{hubbard,straton}, which is based on the identity
\begin{equation}
\label{hubstrat1}
e^{\beta \hat{\rho}^2} = \frac{1}{\sqrt{2 \pi}} \int_{- \infty}^{\infty} 
        e^{-\frac{\sigma^2}{2}} e^{\sigma \sqrt{2 \beta} \hat{\rho}} d \sigma,
\end{equation}
where $\hat{\rho}$ is a one-body operator.
In order to avoid problems due to non-commuting operators, one can split 
up the Boltzmann operator using the Suzuki-Trotter formula \cite{suzuki}.
One can discretize the Hubbard-Stratonovich transformation
by applying a Gaussian quadrature formula to the integral over $\sigma$.
After a Suzuki-Trotter expansion in $N_t$ slices, 
a three-points quadrature formula leads to an error of the order of 
${\cal O} \left( \frac{\beta^3}{N_t^2} \hat{h}^3\right)$.
This is of the same order in $\beta$ and $N_t$ 
as the errors due to the non-commutativity of the squared operators
that build up $\hat{h}$. 
For some systems one can derive an exact,
discrete Hubbard-Stratonovich transform. 
Hirsch showed how one can write an operator of the form 
$e^{-\beta U \hat{n}_{\uparrow} \hat{n}_{\downarrow}}$,
where $\hat{n}_{\sigma}$ is the site occupation number for an
electron with spin projection $\sigma$,  exactly as a sum of two 
exponentials of one-body operators \cite{hirsch}.
Recently, Gunnarsson and Koch extended this to systems 
with higher orbital degeneracy \cite{gunar}.

The aim of this paper is to describe another discrete decomposition scheme,
which is exact for a certain class of operators.
This decomposition scheme is generalized to any two-body Hamiltonian 
using the Suzuki-Trotter formula. 
For the application in AFQMC methods, especially the shell-model 
Quantum Monte-Carlo method, this new decomposition has the advantage,
compared to the discretized Hubbard-Stratonovich transform based 
on Eq.(\ref{hubstrat1}), that it is more accurate and that it leads to 
low-rank matrices. This leads to faster matrix multiplications
and requires less computer memory.
AFQMC methods for fermions often have sign problems \cite{lind}. 
Fahy and Hamann \cite{fahy} showed that these sign problems can be related to
the diffusive behavior of states in the Hubbard-Stratonovitch transformation. 
Because our decomposition, in general, is based on exponentials of one-body 
operators of a completely different type, 
one can expect different sign properties.
Our decomposition is not free of sign problems, but there might be systems
where it leads to a sign rule while the 
Hubbard-Stratonovich transformation does not, 
or where our decomposition causes significally less sign problems.

In Section II we introduce a matrix notation for Slater determinants 
and operators needed for a clear discussion of the decomposition.
In Section III a basic lemma is given on which the decomposition is based.
In Section IV the exact decomposition for a certain class of operators 
is presented. We indicate how to apply this decomposition to a
general two-body Hamiltonian.
In Section V the relation with Hirsch's decomposition for the Hubbard model
is elucidated.
Finally, in Section VI the decomposition for the nuclear pairing Hamiltonian 
is discussed and illustrated with AFQMC-results for an exactly solvable model.

\section{A matrix notation for Slater determinants and operators}
In order to avoid confusion between matrix representations in the space
of single-particle states and the operators themselves in Fock space, 
we will denote the former with upper case
and the latter with lower case characters.
Let $\left\{ \phi_1,\ldots,\phi_{N_s} \right\}$ be a set of basis states
for the single-particle space,
$\hat{a}_1, \ldots , \hat{a}_{N_s}$ be the  related creation operators and
the $A$-particle state $ \Psi_M$ the antisymmetrized product of a set of 
single-particle states $\sum_{i=1}^{N_s} M_{i j} \phi_i, \ j=1 \ldots A$.
i.e.  $\Psi_M$ is a Slater determinant.
Thus in second quantization one can write
\begin{equation}
 | \Psi \rangle = \prod_{j=1}^A 
        \left( \sum_{i=1}^{N_s} M_{i j} \hat{a}_i \right) | \rangle.
\end{equation}
This defines a matrix representation $M$ for a Slater determinant $\Psi_M$.
The value of this representation is that one can represent 
certain operations on the Slater determinant by matrix operations on M.
e.g. the overlap between two Slater determinants $\Psi_{M'}$ and $\Psi_{M}$
is given by
$ \langle \Psi_{M} | \Psi_{M'} \rangle = \det \left(M^{\dagger} M' \right)$. 
The exponential of a one-body operator acting on $\Psi_M$
results in a new Slater determinant,
$e^{-\beta \hat{h}} | \Psi_M \rangle = | \Psi_{M'} \rangle$
(this is a corollary of the Thouless-theorem \cite{thouless}),
whose matrix representation is related to $M$ by
$ M'= e^{-\beta H} M $,
where the $N_s \times N_s$ matrix  $H$ is defined by
$ \hat{h} = \sum_{i,j} H_{i j} \hat{a}_i^{\dagger} \hat{a}_j$.
Reversily, given a $N_s \times N_s$ matix $Q$, one can consider 
the operator $\hat{\cal O}(Q)$, defined by its action on Slater determinants:
\begin{equation}
	\hat{\cal O }(Q): \ \ | \Psi_M \rangle \ \longrightarrow \
       \hat{\cal O }(Q)  | \Psi_M \rangle = | \Psi_{M'} \rangle
       \ \ \mbox{with} \ M'=Q M .
\end{equation}
If $Q$ is non-singular, $\hat{\cal O}(Q)$ is the exponential 
of a one-body operator.

\section{A basic lemma}
\begin{lemma}
The operation represented by the unit matrix plus a matrix of rank two 
can be expressed as a sum of one- and two-body operators  
in the following way:
\begin{equation}
 \label{r2operator}
 \hat{\cal O}(1+\alpha \, B_1^{\dagger} B_4 + \beta \,  B_2^{\dagger} B_3 ) 
  =  1 + \alpha \, \hat{b}_1^{\dagger} \hat{b}_4 
       +\beta \, \hat{b}_2^{\dagger} \hat{b}_3
       + \alpha \beta \,  \hat{b}_1^{\dagger}  \hat{b}_2^{\dagger} 
                         \hat{b}_3 \hat{b}_4, 
\end{equation}
where $B_1, B_2, B_3$ and $B_4$ are $1 \times N_s$ row matrices and
$ \hat{b}_k = \sum_{j=1}^N (B_k)_j \, \hat{a}_j$ ,  $k=1,2,3,4$.
\end{lemma}
\begin{proof}
Consider the $A$-particle Slater determinant $\Psi_M$ 
represented by the matrix $M$.
Consider also a Slater determinant $\Psi_L$, that has particles in the 
single-particle states $\phi_{i_1},\phi_{i_2},\ldots,\phi_{i_A}$.
The Slater determinants of this type constitue a basis of the $A$-particle
Hilbert space.
The overlap of $\Psi_L$ with $\Psi_M$ is given by
\begin{equation}
\label{r2over}
\langle \Psi_L | \Psi_M \rangle = \det \left(
\tilde{M}_{. 1} \ \tilde{M}_{. 2} \ \cdots \ \tilde{M}_{. A}
\right).
\end{equation}
The notation $M_{. j}$ denotes the vector that is given 
by the $j^{th}$ column of $M$,
the notation $\tilde{B}$ for an $N$-element vector $B$ denotes the $A$-element 
vector $( B_{i_1}   B_{i_2}  \cdots  B_{i_A} )$.  
The operator in Eq.(\ref{r2operator}) transforms $\Psi_M$ into $\Psi_{M'}$ 
with $M'= (1 + \alpha B_1^{\dagger} B_4 + \beta B_2^{\dagger} B_3) M$.
To calculate the overlap of $\Psi_{M'}$ with $\Psi_L$,
we have to replace every column $\tilde{M}_{. j}$ in Eq.(\ref{r2over}):
\begin{equation}
  \tilde{M}_{. j}  \;  \rightarrow \; \tilde{M}'_{. j} 
  = \tilde{M}_{. j} + \alpha_j \tilde{B}_1^{\dagger}
                    + \beta_j \tilde{B}_2^{\dagger},
    \ \ \mbox{with} \
    \alpha_j= \alpha \ B_4 M_{. j},  \ \ 
    \beta_j = \beta  \ B_3 M_{. j}.
\end{equation}
The overlap is then given by
\begin{equation}
\label{r2over2}
\langle \Psi_L | \Psi_{M'} \rangle =
  \det \left(
\tilde{M}_{. 1} + \alpha_1 \tilde{B}_1^{\dagger} 
                + \beta_1 \tilde{B}_2^{\dagger} 
 \ \ \ \cdots \ \ \ 
\tilde{M}_{. A} + \alpha_A \tilde{B}_1^{\dagger} 
                + \beta_A \tilde{B}_2^{\dagger} 
\right).  
\end{equation}
This determinant can be expanded as the sum of all determinants 
that are obtained by selecting in every column of Eq.(\ref{r2over2}) 
one of the terms $\tilde{M}_{. j}$, $ \alpha_j \tilde{B}_1^{\dagger}$ or 
$\beta_j \tilde{B}_2^{\dagger}$.
If in more than one column the term $ \alpha_j \tilde{B}_1^{\dagger}$ 
is selected, then the determinant has two linearly dependent columns, 
so it will vanish.
The same holds for the term $\beta_j \tilde{B}_2^{\dagger}$.
Only four types of determinants remain:
\begin{itemize}
\item
  $\tilde{M}_{.}$ is selected in every column. 
  This determinant is just $\langle \Psi_L | \Psi_M \rangle $
  (see Eq.(\ref{r2over})).
\item 
  $ \alpha_j \tilde{B}_1^{\dagger}$ is selected in column $j$, 
  $\tilde{M}_{.}$ in all others. These determinants sum up to  
  $\langle \Psi_L |\alpha \hat{b}_1^{\dagger} \hat{b}_4 | \Psi_M \rangle$
  (one particle is moved from state $b_4$ to state $b_1$).
\item 
  $ \beta_j \tilde{B}_2^{\dagger}$ is selected in column $j$, 
  $\tilde{M}_{.}$ in all others. These determinants sum up to  
  $\langle \Psi_L |\beta \hat{b}_2^{\dagger} \hat{b}_3 | \Psi_M \rangle$
  (one particle is moved from state $b_3$ to state $b_2$).
\item
  $ \alpha_j \tilde{B}_1^{\dagger}$ is selected in column $j$, 
  $ \beta_k \tilde{B}_2^{\dagger}$ is selected in column $k$, 
  $\tilde{M}_{.}$ in all others. These determinants sum up to 
  $\langle \Psi_L |\alpha \beta \hat{b}_1^{\dagger} \hat{b}_2^{\dagger} 
  \hat{b}_3 \hat{b}_4 | \Psi_M \rangle$
  (two particles are moved from states $b_4$ and $b_3$ 
   to states $b_1$ and $b_2$).
\end{itemize}
Taking all these terms together, we find that
\begin{equation}
\langle \Psi_L | \Psi_{M'} \rangle = \langle \Psi_L |
1+\alpha\hat{b}_1^{\dagger}\hat{b}_4 + \beta\hat{b}_2^{\dagger}\hat{b}_3
+\alpha\beta \hat{b}_1^{\dagger}\hat{b}_2^{\dagger}\hat{b}_3\hat{b}_4
| \Psi_{M} \rangle .
\end{equation}
This holds for any basis state $\Psi_L$, so that
\begin{equation}
 \Psi_{M'} =
\left( 1+\alpha\hat{b}_1^{\dagger}\hat{b}_4 + \beta\hat{b}_2^{\dagger}\hat{b}_3
+\alpha\beta \hat{b}_1^{\dagger}\hat{b}_2^{\dagger}\hat{b}_3\hat{b}_4 \right)
\Psi_{M}.
\end{equation}
This proves Eq.(\ref{r2operator}).
\end{proof}

\section{A discrete Hubabrd Stratonovich decomposition}
Consider a fermionic two-body operator $\hat{Q}$ of the form
\begin{equation}
\label{exdecoform}
\hat{q} = \sum_{i,j,k,l=1}^{N_s} 
               Q_{i j} \left(B_1\right)_k \left(B_2\right)_l \, 
               \hat{a}^{\dagger}_i \hat{a}^{\dagger}_j \hat{a}_l \hat{a}_k.
\end{equation}
An operator of this form has the special property that
\begin{equation}
   \hat{q}^2=\lambda \hat{q},
   \ \ \mbox{with} \
   \lambda=\sum_{k,l=1}^{N_s} \left( Q_{k l} - Q_{l k} \right) 
                       \left(B_1\right)_k \left(B_2\right)_l.
\end{equation}
Because of this relation, the exponential of $\hat{q}$ can be written as
\begin{equation} 
   e^{-\beta \hat{q}} = 1 + \gamma \hat{q}, 
   \ \ \mbox{with} \ 
   \left\{ \begin{array}{ccccc}
   \gamma &=& \frac{ e^{-\beta \lambda} -1 } {\lambda} 
     & {\rm for} & \lambda \neq 0, \\
   \gamma &=& -\beta   
     & {\rm if} & \lambda = 0.
\end{array} \right.
\end{equation}
Now we can use the lemma to obtain a discrete decomposition of
$e^{-\beta \hat{q}}$ in a sum of exponentials of one-body operators:
\begin{eqnarray}
  \nonumber
  e^{-\beta \hat{q}} &=& 
    \sum_{i,j=1}^{N_s} \frac{1}{2} \sum_{\sigma=-1,+1} 
         \frac{ |Q_{i j}|}{\Theta}  \, \left( 1 
      +  \sigma \chi_{i j}  \hat{a}^{\dagger}_i\hat{b}_1
      +  \sigma \chi'_{i j} \hat{a}^{\dagger}_j \hat{b}_2
      +  \chi_{i j} \chi'_{i j} 
            \hat{a}^{\dagger}_i \hat{a}^{\dagger}_j \hat{b}_2 \hat{b}_1
	\right), 
 \\ \label{exdisco}
 & = & \sum_{i,j=1}^{N_s} \sum_{\sigma=-1,+1} \frac{ |Q_{i j}|}{2 \Theta} \,
         \hat{\cal O} \left( 1
      +  \sigma \chi_{i j}  A_i^{\dagger} B_1
      +  \sigma \chi'_{i j} A_j^{\dagger} B_2
	\right), 
\end{eqnarray}
with $A_i$ the $1 \times N_s$ row matrix which has a 1 on the $i^{th}$ entry
and zeros anywhere else, and 
\begin{eqnarray}
 \Theta      & = & \sum_{i,j=1}^{N_s} |Q_{i j}|, \\
 \chi_{i j}  & = & \sqrt{|\gamma| \Theta }, \\
 \chi'_{i j} & = & \sqrt{|\gamma| \Theta } \,
                         \mbox{sign} \left(\gamma Q_{i j}\right), \\
 \hat{b}_k   & = & \sum_{l=1}^{N_s} \left(B_k\right)_l \hat{a}_l, \ \ k=1,2. 
\end{eqnarray}
This is an exact Hubbard-Stratonovich-like decomposition
of the form of Eq.(\ref{exdecoform}). 
To apply this discrete Hubbard-Stratonovich-like decomposition to the 
Boltzmann operator $e^{-\beta \hat{v}}$ for a general fermionic  
two-body operator $\hat{v}$, one has to rewrite  $\hat{v}$ as a sum of
operators of the form Eq.(\ref{exdecoform}).
A trivial way to do this, is given by
\begin{equation} 
   \hat{v}= \sum_{i,j,k,l=1}^{N_s} V_{i j k l} 
              \hat{a}^{\dagger}_i \hat{a}^{\dagger}_j \hat{a}_l \hat{a}_k
	  = \sum_{k,l=1}^{N_s} \hat{q}_{k l},
   \ \ \mbox{with} \ 
   \hat{q}_{k l}=\left( \sum_{i,j=1}^{N_s}  V_{i j k l} 
              \hat{a}^{\dagger}_i \hat{a}^{\dagger}_j \right) 
              \,  \hat{a}_l \hat{a}_k.
\end{equation}
The Suzuki-Trotter formula can be used to split up
the Boltzmann operator into factors with only one operator $\hat{q}_{k l}$
in the exponent. 
Then each of these factors can be decomposed exactly using the discrete
decomposition in Eq.(\ref{exdisco}).
Note that the total decomposition is no longer exact because of the
non-commutativity of the operators $\hat{q}_{k l}$.
The error will be of the order ${\cal O} (\beta^3 / {N_t}^2)$.
It will be much smaller than in case of a decomposition based on a 
Gaussian discretization of the integral in Eq.(\ref{hubstrat1}),
because now the error is proportional to the commutators of the operators
$\hat{q}_{k l}$ and no longer to a power of $\hat{v}$.

\section{Relation to Hirsch's decomposition for the Hubbard Hamiltonian}
For the Hubbard model we have to find a decomposition for a Boltzmann operator
of the form $e^{-\beta  U  \hat{n}_{\uparrow}  \hat{n}_{\downarrow}}$.
where $U$ is the interaction strength and
$\hat{n}_{i \sigma} = \hat{a}^{\dagger}_{\sigma} \hat{a}_{\sigma}$.
$\sigma = \uparrow, \downarrow$ is an index for the spin degree-of-freedom.
The exponent has a two-body operator 
$\hat{n}_{\uparrow}  \hat{n}_{\downarrow}$, 
which is an operator of the form of Eq.(\ref{exdecoform}),
so we can apply the decomposition given in Eq.(\ref{exdisco}) and obtain:
\begin{equation}
\label{hirschdeco}
 e^{-\beta  U  \hat{n}_{\uparrow}  \hat{n}_{\downarrow}} =
       \frac{1}{2} \sum_{\sigma=-1,+1} \hat{\cal O} \left( 
       1+ \sigma \chi_{ \uparrow} N_{ \uparrow} 
        + \sigma \chi_{\downarrow} N_{\downarrow} \right),
\end{equation}
with $N_{ \uparrow} (N_{ \downarrow})$ the matrix 
which is zero everywhere except for the diagonal element related 
to the spin-up (spin-down) site, which is equal to 1.
$\chi$ and $\chi'$ are given by 
\begin{eqnarray} 
  & \chi_{ \uparrow} = -\chi_{\downarrow} =  \sqrt{1-e^{-\beta U}} &
  \ \ \mbox{for} \ \beta U > 0, \\
  \mbox{or} \ \ &
   \chi_{ \uparrow} = \chi_{\downarrow} =  \sqrt{e^{-\beta U} -1} &
  \ \ \mbox{for} \ \beta U < 0.
\end{eqnarray}
Now one could scale each term in Eq.(\ref{hirschdeco}) 
with an operator of the form 
$e^{-\beta \mu \left(\hat{n}_{\uparrow} + \hat{n}_{\downarrow}\right)}$,
because in the canonical ensemble this is just a constant.
The matrices in the decomposition now have to be multiplied with the matrix
$1+ \left(e^{-\beta \mu}-1\right) N_{\uparrow} 
  + \left(e^{-\beta \mu}-1\right) N_{\downarrow}$.
In case of the repulsive Hubbard model, the choice $\mu=-U/2$ 
leads to the discrete Hubbard-Stratonovich transform of Hirsch \cite{hirsch}.
From the computational point of view this particular choice of $\mu$
has the advantage that the matrix representation for the spin-down part
is related to the matrix representation for the spin-up part by a matrix 
inversion. Then one only has to keep track of the spin-up part in actual 
AFQMC calculations.  
Hirsch's decomposition for the attractive Hubbard model can also be obtained
from Eq.(\ref{hirschdeco}), with a particular choice for $\mu$. 
In this case however, there is no computational 
advantage in taking any particular value.
\section{Application to the nuclear pairing Hamiltonian}
The Hamiltonian for nuclear pairing in a degenerate shell is given by
\begin{equation}
   \hat{h}= - G \sum_{k,k'=1}^{N_S} 
           \hat{a}^{\dagger}_k \hat{a}^{\dagger}_{\bar{k}}
           \hat{a}_{\bar{k}'} \hat{a}_{k'}.
\end{equation}
Here it is assumed that there are $2 N_S$ degenerate single-particle states.
The single-particle energy is shifted to $0$ MeV.
So there is no one-body part in the Hamiltonian.
The states with $j_z>0$ are labeled from 1 to $N_S$ and 
$\bar{k}$ denotes the time-reversed state of state $k$.
The many-body problem for this model can be solved analytically
using the seniority scheme \cite{ring}.

Using the Suzuki-Trotter formula,
the Boltmann operator for this Hamiltonian can be written as
\begin{equation}
 e^{-\beta \hat{h}}=
 e^{-\frac{\beta}{2} \hat{q}_1} \,  
 e^{-\frac{\beta}{2} \hat{q}_2} \, \cdots \,
 e^{-\frac{\beta}{2} \hat{q}_{N_S} } \, 
 e^{-\frac{\beta}{2} \hat{q}_{N_S} } \,  \cdots \,
 e^{-\frac{\beta}{2} \hat{q}_2} \,  
 e^{-\frac{\beta}{2} \hat{q}_1}
 \ + \, {\cal O} \left( \beta^3 \right),
\end{equation}
with
\begin{equation}
 \hat{q}_k = - G  \left( \sum_{k'=1}^{N_S} 
           \hat{a}^{\dagger}_{k'} \hat{a}^{\dagger}_{\bar{k'}} \right)
           \hat{a}_{\bar{k}} \hat{a}_{k}.
\end{equation}
The error is of the order ${\cal O} \left(\beta^3\right)$.
It is assumed that $\beta$ is small. 
In practice, one has to split $\beta$ in a number of 
inverse-temperature slices using the Suzuki-Trotter formula.
Then one can apply the procedure that is discussed here
to each inverse-temperature slice seperately. 
We have $\hat{q}_k^2=-G \hat{q}_k$. 
So we can find a decomposotion of the type given in Eq.(\ref{exdisco})
\begin{equation}
\label{pairdeco}
  e^{-\frac{\beta}{2} \hat{q}_k} = 
	\sum_{k'=1}^{N_S} \sum_{\sigma=-1,+1} \frac{1}{2 N_S} \,
         \hat{\cal O} \left( 1
      +  \sigma \chi  A_{k'}^{\dagger} A_k
      +  \sigma \chi  A_{\bar{k}'}^{\dagger} A_{\bar{k}}
	\right), 
\end{equation}
where $\chi^2 = N_S \left( e^{\frac{\beta G} {2}} -1 \right)$ .
 
\begin{figure}
\epsfig{file=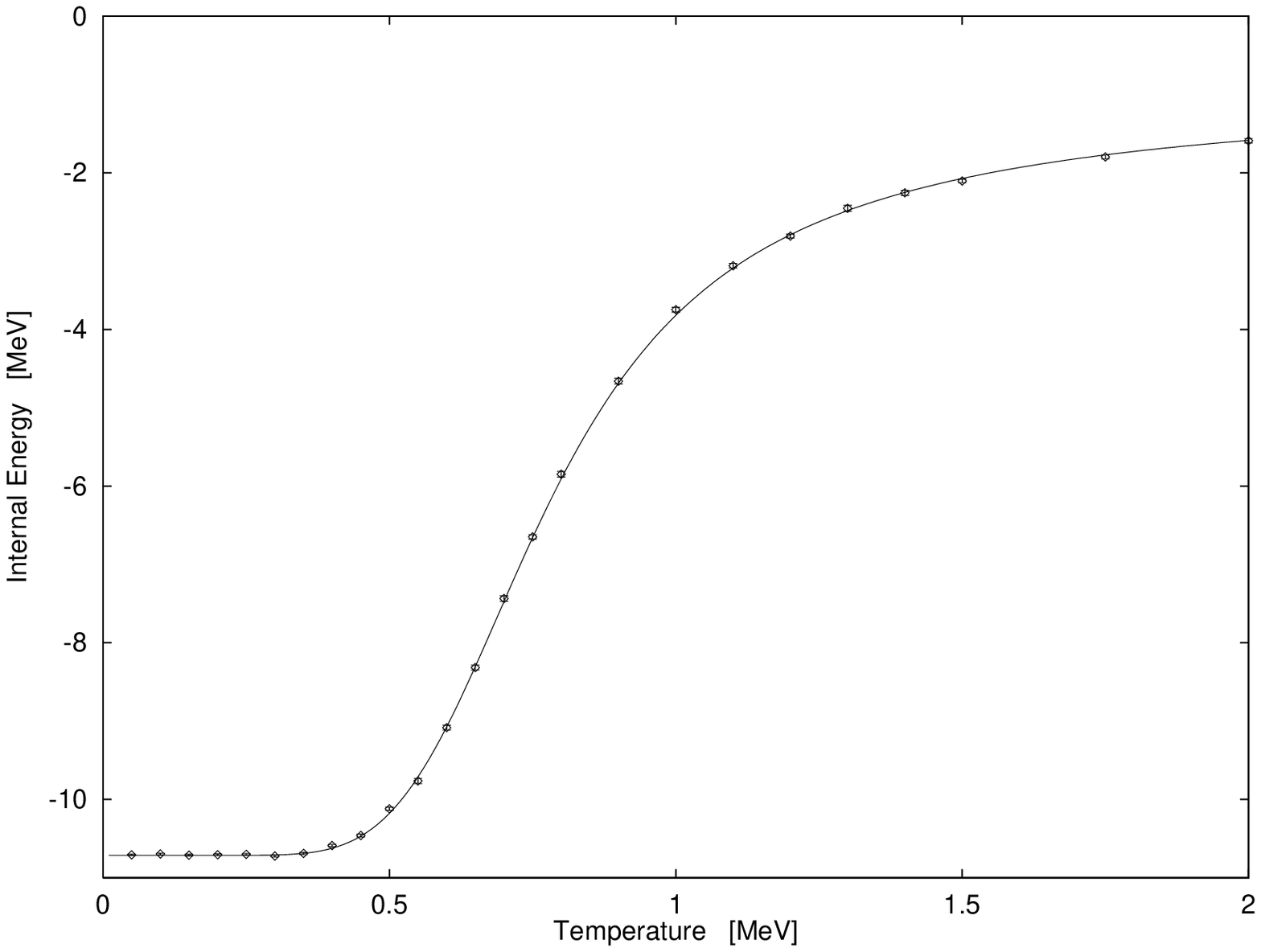,height=9cm}
\caption{Internal energy versus temperature. 
Error bars on the Monte-Carlo data are omitted 
because they are smaller than the symbols marking the data points.
}
\label{fig1}
\end{figure}

\begin{figure}
\epsfig{file=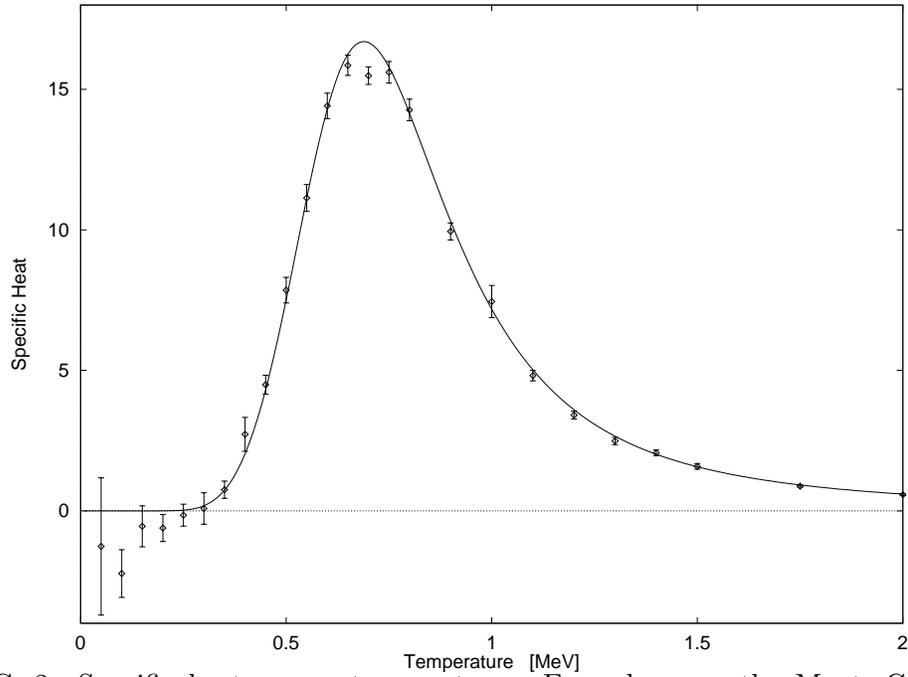,height=9cm}
\caption{Specific heat versus temperature. 
Error bars on the Monte-Carlo data represent 95\%-confidence intervals. 
}
\label{fig2}
\end{figure}

We have applied this decomposition to study a degenerate shell of 20 states 
$(N_S=10)$, with 10 particles.
This could model the valence model space for neutrons in the $fp$ shell 
in $^{56}\mbox{Fe}$,
if one neglects the energy gap between the $1f_{\frac{7}{2}}$ and the 
$2p_{\frac{3}{2}}$, $1f_{\frac{5}{2}}$, $2p_{\frac{1}{2}}$ orbitals.
For the strength of the interaction we took 
$G=20/A$ MeV $=20/56$ MeV, as recommended in \cite{bes}.
We have performed a shell-model quantum Monte-Carlo calculation 
in the canonical ensemble, following \cite{smmc}, 
but now using the new decomposition of Eq.(\ref{pairdeco})
instead of the Hubbard-Stratonovich transformation.
In order to make the systematic error smaller than the statistical error,
the inverse temperature $\beta$ was split into slices of length 
$0.05 \ \mbox{MeV}^{-1}$.
We point out that the form
$1+ \sigma \chi A_{k'}^{\dagger} A_k 
  + \sigma \chi  A_{\bar{k}'}^{\dagger} A_{\bar{k}} $
can be rewritten as  $\left(1+ \sigma \chi A_{k'}^{\dagger} A_k \right) \,
 \left(1+ \sigma \chi  A_{\bar{k}'}^{\dagger} A_{\bar{k}}\right) $,
such that there is a symmetry between states with $j_z>0$ and their
time-reversed states. 
This symmetry guarantees that there will be no sign problem for systems
with an even number of particles.
This sign-rule is analogous to the sign rule for 
the pairing-plus-quadrupole Hamiltonian 
decomposed using the Hubbard-Stratonovich transform \cite{lang}.
In figure \ref{fig1} we show the internal energy of the system as a function of
temperature.
In figure \ref{fig2} we show the corresponding specific heat of the system 
as a function of temperature. 
The Monte-Carlo results are in excellent agreement with the analytical results.
The peak in de specific-heat curve around a temperature of $0.8$ MeV.
can be associated with the breakup of $J^\pi=0^{+}$ pairs.
It is straightforward to take into account the different single-particles
energies and more general forms of the pairing Hamiltonian:
\begin{equation}
 \hat{h}= 
    \sum_{k=1}^{N_S} \epsilon_k \left(\hat{n}_k + \hat{n}_{\bar{k}} \right)
 \, - \,   \sum_{k,k'=1}^{N_S} G_{k ,k'} 
           \hat{a}^{\dagger}_k \hat{a}^{\dagger}_{\bar{k}}
           \hat{a}_{\bar{k}'} \hat{a}_{k'}.
\end{equation}
Extension to even more general two-body Hamiltonians is possible.
Then there can be sign problems at low temperatures.

\section{Conclusion}
We have presented a new type of discrete Hubbard-Stratonovich decomposition 
for the Boltzmann operator. 
It is exact for a special class of two-body operators.
Applied to the Hubbard Hamiltonian, it leads to Hisrch's discrete  
Hubbard-Stratonovich decomposition.
The decomposition is well suited for the nuclear pairing Hamiltonian,
where it leads to a sign rule for systems with an even number of particles. 
Quantum Monte-Carlo results based on this decomposition
are in excellent agreement with the analytical results 
for an exactly solvable model.

\section*{Acknowledgements}
The authors are grateful to the F.W.O. (Fund for Scientific Research) 
- Flanders and to the Research Board of the University of Gent for
financial support.

\end{document}